\documentclass{article}
\begin{document}
\title{Einstein-Cartan cosmology and the CMB anisotropies}
\author{Davor Palle \\
ul. Ljudevita Gaja 35, 10000 Zagreb, Croatia \\
email: davor.palle@gmail.com}
\maketitle
\begin{abstract}
{We derive linear scalar perturbation equations for Einstein-Cartan field equations
of Weyssenhoff fluid, as well as for the corresponding perturbations of
Bianchi identity and geodesic equations. The equations are given in
both conformal Newtonian and synchronous gauges. They are suitable for
numerical implementation when precise evolution of torsion and its
perturbation will be extracted from N-body cosmic simulations of the large
scale structures in the Universe. A rising number of problems of the
concordance cosmological model forces us to include the rotational
degrees of freedom realized through torsion in the Einstein-Cartan gravity.}
\end{abstract}

\section{Introduction and motivation}
Despite the huge success of the concordance $\Lambda CDM$ model, cosmologists
are faced with new challenges of the theory rooted in the surprising
observational results known as the Hubble tension, 21cm EDGES anomalous
absorption signal or anomalous high-redshift galaxy halo number densities
(see \cite{Palle1} and references therein).
It seems that some of the above problems could be resolved within 
the Einstein-Cartan cosmology \cite{Palle1}.

Since there are a large number of cosmic observables defined as
small perturbations of various physical quantities, we derive 
in this paper scalar perturbations for the CMB, neutrinos, baryons
and CDM within the Einstein-Cartan cosmology.

A detailed description of the framework and all equations can be found
in the next chapter and in Appendix A. The last chapter and the Appendix B
contain some comments and suggestions for numerical implementation.

\section{Linear scalar perturbation equations}
We follow closely the definitions of ref. \cite{Ma} of spatially flat
cosmology with metric assignment (-+++) and the standard relation between the
proper t and conformal $\tau$ time $d\tau = dt/a(\tau)$, while derivatives
are denoted by dots: $\dot{a}\equiv \partial a/\partial \tau$.

All equations will be given in the Fourier k-space with the following
definition for any G:
\begin{eqnarray*}
G(\vec{x},\tau) = \int d^{3}k e^{\imath \vec{k}\cdot\vec{x}} G(\vec{k},\tau).
\end{eqnarray*}

The Einstein-Cartan cosmological model \cite{Palle2,Palle3,Palle1} is
spatially flat with $\Omega_{tot}=\Omega_{m}+\Omega_{Q}+\Omega_{\Lambda}=1$, but
with very well determinate $\Omega_{m}=2,\ \Omega_{Q}=-1\ and\ \Omega_{\Lambda}=0$.
The gauge invariant perturbed densities in the spacetimes with vorticity or shear
have more complex structure \cite{Palle3} than in Friedmann spacetimes.
Fortunately, the observations suggest that we can ignore small deviations
from isotropy and homogeneity of Friedmann geometry. Thus, we perform perturbations
in the conformal Newtonian and synchronous gauges on the Friedmann background \cite{Ma}
with the effective energy-momentum tensor of the Einstein-Cartan (EC) Weyssenhoff fluid
model \cite{Obukhov}. It is possible to define an effective 
energy-momentum tensor for any model in the Einstein-Cartan theory of gravity \cite{Hehl}.

Appendix A is dedicated to the detailed definitions and discussion of the EC field 
equations for Weyssenhoff fluid and the corresponding Bianchi identity.

Acknowledging the relations of Appendix A and flat geometry perturbation theory 
\cite{Kodama,Ma} we arrive at the perturbed EC field equations in the conformal Newtonian
gauge (equations analogous to eq. (23a)-(23d) of ref.\cite{Ma}; $\phi$ and $\psi$
are metric perturbations):
\begin{eqnarray}
&&3\frac{\dot{a}}{a}(\dot{\phi}+\frac{\dot{a}}{a}\psi)+k^2\phi=-\frac{a^2}{2}\kappa\delta \rho
+a^2 Q \delta Q, \nonumber \\
&&k^2(\dot{\phi}+\frac{\dot{a}}{a}\psi)=\frac{a^2}{2}\kappa(\rho+p)\imath\vec{k}\cdot\vec{v}
-a^2 Q^2 \imath\vec{k}\cdot\vec{v}, \nonumber \\
&&\ddot{\phi}+\frac{\dot{a}}{a}(\dot{\psi}+2 \dot{\phi})+(2\frac{\ddot{a}}{a}
-\frac{\dot{a}^2}{a^2}) \psi +\frac{k^2}{3}(\phi-\psi)=\frac{1}{2}\kappa a^2\delta p
-a^2 Q\delta Q, \nonumber \\
&&k^2(\phi - \psi)=12\pi G_{N}a^2(\rho +p)\sigma.
\end{eqnarray}

The effective energy-momentum tensor appears as:
\begin{eqnarray}
T_{\mu\nu}^{eff}&=&(p-\kappa S^2-\Lambda)g_{\mu\nu}+U_{\mu}U_{\nu}(p+\rho-2\kappa S^2) \nonumber \\
& &-2(-g^{\alpha\beta}+U^{\alpha}U^{\beta})\nabla_{\alpha} [U_{(\mu}S_{\nu)\beta}],   \\
&&only\ Q_{12}=-Q_{21}\neq 0,\ Q^2=\frac{1}{2}Q_{\mu\nu}Q^{\mu\nu},\ 
\kappa=8\pi G_{N},\ Q=\kappa S. \nonumber
\end{eqnarray}

We apply the assumption of isotropy (Friedmann geometry) deriving the above equations neglecting the
terms proportional to torsion such as $k_{2}v_{1}-k_{1}v_{2}$ or $k_{1}k_{2}(v_{1}-v_{2})$.
Perturbations of the EC field equations in the synnchronous gauge contain the same torsion terms
as in the conformal Newtonian gauge.

Perturbation of the EC Bianchi identities described in Appendix A leads us to
the following equations for CDM and baryon density contrasts and velocity gradients in 
the conformal Newtonian gauge:
\begin{eqnarray}
\dot{\delta_{c}}&=& -\Theta_{c} + 3\dot{\phi}, \nonumber \\
\dot{\Theta_{c}}&=&[1+(-7+\frac{1}{3})\frac{Q^2}{\kappa\rho_{c}}]^{-1}\{-\frac{\dot{a}}{a}\Theta_c 
+k^2\psi \nonumber \\
&-&\frac{1}{\kappa\rho_{c}}[2k^2 Q\delta Q+k^2 (5-\frac{1}{3})Q^2\psi
-8Q\dot{Q}\Theta_{c}-12Q^2\frac{\dot{a}}{a}\Theta_{c}]\}, \nonumber \\
\dot{\delta_{b}}&=& -\Theta_{b} + 3\dot{\phi}, \nonumber \\
\dot{\Theta_{b}}&=&[1+(-7+\frac{1}{3})\frac{Q^2}{\kappa\rho_{b}}]^{-1}\{-\frac{\dot{a}}{a}\Theta_b 
+k^2\psi +c_{s}^{2}k^2 \delta_{b} \nonumber \\
&-&\frac{1}{\kappa\rho_{b}}[2k^2 Q\delta Q+k^2 (5-\frac{1}{3})Q^2\psi
-8Q\dot{Q}\Theta_{b}-12Q^2\frac{\dot{a}}{a}\Theta_{b}] \nonumber  \\
&+&\frac{4\rho_{\gamma}}{3\rho_{b}}an_{e}x_{e}\sigma_{T}(\Theta_{\gamma}-\Theta_{b})\}, \\
\Theta &\equiv& \imath \vec{k}\cdot\vec{v}. \nonumber
\end{eqnarray}

Density contrasts in the synchronous gauge do not contain torsion terms, just like in the 
Newtonian gauge. $\dot{\Theta_{c}}(synch)$ vanishes, while $\dot{\Theta_{b}}(synch)$ has the same
form as $\dot{\Theta_{b}}(conf)$, but without terms proportional to $\psi$.
We discard terms that should vanish owing to the isotropy and put $\iota k_{3}v_{3}
=\frac{1}{3}\iota \vec{k}\cdot\vec{v}$ using the same argument.

The Boltzmann equations for the phase-space distributions require the resolution of
the perturbed geodesic equations in the EC cosmology:
\begin{eqnarray*}
&&P^{0}\frac{P^{\mu}}{d\tau}+\tilde{\Gamma}^{\mu}_{(\nu\kappa)}P^{\nu}P^{\kappa}=0, \\
&&\tilde{\Gamma}^{\mu}_{(\nu\kappa)}=\left\{ \begin{array}{c} \mu \\ \nu\kappa \end{array}\right\}+ Q^{\ \ \mu}_{\nu\kappa .}+Q^{\ \ \mu}_{\kappa\nu .},\   
(\mu\nu)=\frac{1}{2}(\mu\nu+\nu\mu),  \\
&&\tilde{\Gamma}^{\mu}_{\nu\kappa}=\left\{ \begin{array}{c} \mu \\ \nu\kappa \end{array}\right\}+Q^{\     \ \mu}_{\nu\kappa .}
+Q^{\ \ \mu}_{\kappa\nu .}+Q^{\mu}_{. \nu\kappa}, \\ 
&&torsion\ tensor = Q^{\mu}_{. \nu\kappa} = \frac{1}{2}(\tilde{\Gamma}^{\mu}_{\nu\kappa}-\tilde{\Gamma}^{\mu}_{\kappa\nu}).
\end{eqnarray*}

We verify that the torsion terms cancel out in the perturbed geodesic equations to linear
order in both gauges. As a consequence, the Boltzmann equations for photons, massless
and massive neutrinos retain their forms as in the Einstein cosmology (see ref. \cite{Ma}).

Now we have a complete set of coupled equations for the CDM, baryons, photons and
neutrinos in the EC cosmology.

\section{Conclusion and comments}
Inspecting the form of the perturbation equations in the EC cosmology, one can notice
that we need the knowledge not only of the torsion, but also of its time derivative
and its perturbation (in the Zeldovich model is $\delta Q=Q(\frac{\Omega_{c}}
{\Omega_{m}}\delta_{c}+\frac{\Omega_{b}}{\Omega_{m}}\delta_{b})$).
 We can achieve this objective only with the extensive N-body
numerical simulations within the EC cosmology starting at large redshifts
with a primordial vorticity of the Universe that causes the nonvanishing angular
momentum of the Universe which is a nonrelativistic limit of torsion.
In Appendix B we suggest how to improve the numerical codes for the CMB anisotropy calculations.

Recent analysis of a parity violation in polarization data of Planck \cite{Minami}
refers to the right-handed characteristic.
We show in ref. \cite{Palle4} that the Universe must have a preference to right-handedness
of its vorticity as a consequence of a left-handed weak interactions and the resulting
abundant right-handed helicity light Majorana neutrinos.

The appearance of the primordial cosmic magnetic field is a inevitable consequence
of the existence of the primordial vorticity. All these new phenomena have to
be studied both theoretically and observationally.
\vspace{6mm}

{\bf Appendix A}\newline
In this appendix we adopt metric assignment (+ - - -), as well as all definitions 
and conventions as in ref. \cite{Obukhov} with intention that a reader can 
verify some corrections.

The Riemann-Cartan connection can be expressed as:
\begin{eqnarray*}
\tilde{\Gamma}^{\alpha}_{\beta\mu}=\Gamma^{\alpha}_{\beta\mu}
+Q^{\alpha}_{.\beta\mu}+Q^{\ \ \alpha}_{\beta\mu .}+Q^{\ \ \alpha}_{\mu\beta .}\ .
\end{eqnarray*}

The symmetric part of the EC field equations is:
\begin{eqnarray*}
\tilde{R}_{(\mu\nu)}-\frac{1}{2}\tilde{R}g_{\mu\nu}=\kappa T_{(\mu\nu)}\ .
\end{eqnarray*}

The contracted Bianchi identity has the following form (note the wrong sign
in eqs. (2.15),(4.13),(4.14) and (4.16) of ref. \cite{Obukhov} in front of
the Mathisson-Papapetrou force):
\begin{eqnarray*}
(\tilde{\nabla}_{\nu}-2 Q_{\nu})T^{\nu}_{. \mu}+2Q^{\alpha}_{.\mu\beta}T^{\beta}_{. \alpha}
-S^{\nu}_{.\alpha\beta}\tilde{R}^{\alpha\beta}_{. . \mu\nu}=0 \ .
\end{eqnarray*}

In the equation preceding eq.(5.1) of ref. \cite{Obukhov} the last term is missing:
\begin{eqnarray*}
\tilde{R}_{(\mu\nu)}-\frac{1}{2}\tilde{R}g_{\mu\nu}=R_{\mu\nu}-\frac{1}{2} R g_{\mu\nu}
+2\kappa\nabla_{\alpha}[u_{(\mu}S_{\nu).}^{\ \alpha}] \\
+\kappa^{2}S^{2}(2u_{\mu}u_{\nu}-g_{\mu\nu})
-\kappa g_{\mu\nu}g^{\lambda\phi}\nabla_{\alpha}[u_{(\lambda}S_{\phi).}^{\ \alpha}]\ .
\end{eqnarray*}

However, the effective energy-momentum tensor in eq.(5.2) is correct.

The relation (2.13) of ref. \cite{Obukhov} is not generally fulfilled:
\begin{eqnarray}
(\tilde{\nabla}_{\alpha}-2Q_{\alpha})S^{\alpha}_{. \mu\nu}=T_{[\mu\nu]}\ .
\end{eqnarray}

This is not an obstacle since we have, in general, a free choice to define the energy-momentum
tensor \cite{Hehl}. Anyhow, instead of the tensor in eq. (2.19) of ref. \cite{Obukhov},
we can choose the following one:
\begin{eqnarray*}
T_{\mu\alpha}=(u_{\mu}+z_{\mu})P_{\alpha},\ u^{\mu}z_{\mu}=0,\ z^{\mu}z_{\mu}=-1, \ 
u^{\mu}P_{\mu}=\rho \ .
\end{eqnarray*}

Inserting the above tensor into the eq.(4), we get the algebraic equations for
vector $z^{\mu}$. However, our choice is the "minimal" tensor eq.(5.2) in
ref. \cite{Obukhov}.
\vspace{6mm}

{\bf Appendix B}\newline
A line-of-sight method \cite{Seljak} reduces significantly the time to solve 
the coupled system of equations with photon anisotropies.
Let us write the multipole expansion of the temperature anisotropy:
\begin{eqnarray*}
F_{\gamma}(\tau,k,\mu)=\sum_{l=0}^{\infty}F_{\gamma,l}(\tau,k)
(-\imath)^{l}(2l+1)P_{l}(\mu)\ .
\end{eqnarray*}

It fulfills the following differential equation:
\begin{eqnarray*}
\frac{dF_{\gamma}}{d\tau}+(\imath k\mu+\frac{{d\kappa}}{d\tau})F_{\gamma}=
K_{F}(\phi,\psi,F_{\gamma,0},\Theta_{b},F_{\gamma,2},G_{\gamma,0},G_{\gamma,2}), \\
K_{F}\ is\ well\ known\ function,\ G_{\gamma}\ is\ polarization\ anisotropy.
\end{eqnarray*}

The coupled equations are solved up to some $l_{\gamma}={\cal O}(10)$ and
then the rest of multipoles are evaluated by the line-of-sight integrals \cite{Seljak}
up to some $l_{max}={\cal O}(1000)$.

To avoid the problems with k-sampling and precision, one can instead separate $F_{\gamma}$ 
into the known $\tilde{F}_{\gamma}(l_{\gamma})$ and unknown part 
$\Delta F_{\gamma}(l_{\gamma})$:
\begin{eqnarray*}
F_{\gamma}(\tau,k,\mu)=\tilde{F}_{\gamma}(l_{\gamma},\tau,k,\mu)
+\Delta F_{\gamma}(l_{\gamma},\tau,k,\mu), \\
\tilde{F}_{\gamma}(l_{\gamma},\tau,k,\mu)=\sum_{l=0}^{l_{\gamma}}F_{\gamma,l}(\tau,k)
(-\imath)^{l}(2l+1)P_{l}(\mu), \\
\Delta F_{\gamma}(l_{\gamma},\tau,k,\mu)=\sum_{l=l_{\gamma} +1}^{\infty}F_{\gamma,l}(\tau,k)
(-\imath)^{l}(2l+1)P_{l}(\mu)\ .
\end{eqnarray*}

The unknown part satisfies the differential equation that has an explicit 
solution in the form of integrals:
\begin{eqnarray*}
\frac{d\Delta F_{\gamma}}{d\tau}+(\imath k\mu+\frac{{d\kappa}}{d\tau})\Delta F_{\gamma}=
K_{F}-\frac{d\tilde{F}_{\gamma}}{d\tau}-(\imath k\mu+\frac{{d\kappa}}{d\tau})\tilde{F}_{\gamma}\ .
\end{eqnarray*}

Namely, for known $P(\tau)$ and $R(\tau)$, the equation:
\begin{eqnarray*}
&&\frac{d y(\tau)}{d\tau}+P(\tau) y(\tau) = R(\tau),\ has\ a\ solution: \\
&&y(\tau)=y(\tau_{i})exp[-\int^{\tau}_{\tau_{i}}P(u)d u]
+exp[-\int^{\tau}_{\tau_{i}}P(u)du]\int^{\tau}_{\tau_{i}}dt R(t)
exp[\int^{t}_{\tau_{i}}P(v)dv] \ .
\end{eqnarray*}

It follows then:
\begin{eqnarray*}
F_{\gamma, l}(\tau, k)=\frac{1}{2}(-\imath)^{-l}\int^{+1}_{-1}d\mu\Delta F_{\gamma}(l_{\gamma},
\tau,k,\mu)P_{l}(\mu),\ for\ any\ l > l_{\gamma}\ .
\end{eqnarray*}

Denoting the initial power spectrum with $P_{init}(k)$, the anisotropy spectrum is obtained:
\begin{eqnarray*}
C_{l}(\tau)=N \int d^{3}k P_{init}(k) |F_{\gamma, l}(\tau, k)|^{2} \ .
\end{eqnarray*}

Thus, instead of performing the double derivatives, as in ref.\cite{Seljak}, one has to
evaluate double integrals.

\end{document}